\documentclass[a4paper,11pt]{article}

\usepackage{jheppub}     
\usepackage[T1]{fontenc} 
\usepackage{bm}
\usepackage{mathrsfs}
\usepackage{cancel}
\usepackage[normalem]{ulem}
\usepackage{slashed, braket, dirtytalk, mathtools}
\usepackage{xcolor}
\usepackage{float, soul, parskip, lmodern}
\usepackage{caption}
\usepackage{siunitx}
\usepackage{comment}
\usepackage{subcaption}
\usepackage{gensymb}
\usepackage{natbib}
\usepackage{bbold}
\usepackage{booktabs}

\relax

\def\be{\begin{equation}}
\def\ee{\end{equation}}

\def\sla{\raise.15ex\hbox{$/$}\kern-.57em}

\newcommand{\bear}{\begin{eqnarray}}
\newcommand{\bea}{\begin{eqnarray}}
\newcommand{\bi}{\begin{itemize}}
\newcommand{\ba}{\begin{array}}
\newcommand{\eear}{\end{eqnarray}}
\newcommand{\eea}{\end{eqnarray}}
\newcommand{\ei}{\end{itemize}}
\newcommand{\ea}{\end{array}}

\newbox\pippobox

\def\a{\alpha}		\def\b{\beta}				\def\d{\delta}
						
\def\i{\iota}				\def\l{\lambda}		\def\m{\mu}
						\def\p{\pi}

\def\<{\big\langle}
\def\>{\big\rangle}

\def\simlt{\mathrel{\lower2.5pt\vbox{\lineskip=0pt\baselineskip=0pt
           \hbox{$<$}\hbox{$\sim$}}}}
\def\simgt{\mathrel{\lower2.5pt\vbox{\lineskip=0pt\baselineskip=0pt
           \hbox{$>$}\hbox{$\sim$}}}}

\def\nn{\nonumber}

\title{\boldmath 
Light stringy states and the $g-2$ of the muon}

\preprint{}

\author[a]{Pascal Anastasopoulos,}
\author[a]{Elias Niederwieser}
\author[b,c]{and Fran\c{c}ois Rondeau}

\affiliation[a]{Mathematical Physics Group, Department of Physics, University of Vienna, \\
 Boltzmanngasse 5, 1090 Vienna, Austria.}

\affiliation[b]{Laboratoire de Physique Th\'eorique et Hautes Energies - LPTHE\\ Sorbonne Universit\'e, CNRS, 4 Place Jussieu, 75005 Paris, France.}

\affiliation[c]{Department of Physics, University of Cyprus,\\ Nicosia 1678, Cyprus}

\emailAdd{pascal.anastasopoulos@univie.ac.at}
\emailAdd{a01306455@unet.univie.ac.at}
\emailAdd{frondeau@lpthe.jussieu.fr}


\abstract{In this work, we evaluate the contributions to the anomalous magnetic moment of the muon ($(g-2)_{\mu}$) coming from light stringy states in a D-brane semi-realistic configuration. A scalar which couples only to the muon can have a mass sufficiently light to provide a significant contribution to the $(g-2)_{\mu}$. This scenario can arise in intersecting D-brane models, where such light scalars correspond to the first stringy excitations of an open string stretched between two D-branes intersecting with a very small angle. In this article, we show that there is a region in the space of the geometric parameters of the internal manifold where such scalar light stringy states can explain (part) of the observed discrepancy in the $(g-2)_\m$. In a low string scale framework with $M_s\simgt 10~{\rm TeV}$, we show that an excited Higgs with mass $\mathcal{O}(250~{\rm MeV})$, living in an intersection with an angle of order $\mathcal{O}(10^{-10})$, can provide a significant contribution of one-tenth of the $(g-2)_\m$ discrepancy. This leads to a lower bound for the compact dimension where the branes intersect of order $\mathcal{O}(10^{-8}~{\rm GeV}^{-1})$. We also study patterns in D-brane configurations that realize our proposal, both in three and four stacks models.
}

\begin{document}
\maketitle
\flushbottom

\section{Introduction}
\label{sect:introduction}


Undoubtedly, the formulation of the Dirac equation represents a milestone in developing modern particle physics and the corresponding Standard Model (SM). According to the Dirac theory, the Landé $g$-factor \cite{Land2005} in the expression for the magnetic dipole moment $\vec{\mu}$ of a spin-$\frac{1}{2}$ fermion $f$ with mass $m_f$ and charge $Q_f$, given by \cite{povh1999particles}
\begin{equation}
	\vec{\mu}=g\, \frac{Q_{f}}{2 m_{f}}\,\vec{S} \quad \text{with}\quad \vec{S}=\frac{\vec{\sigma}}{2}~,
\end{equation}
where the $\sigma$'s denote the usual Pauli matrices, could be predicted to the exact value $g=2$. While this remains accurate and true at the tree level, higher loops in the perturbative sense of Quantum Field Theory (QFT) are responsible for deviations of $g$ from $2$.
Generally, the deviation of $g$ from $2$ is measured by the \textit{Pauli form factor} $F_2(p^2)$ at $p^2=0$. Following the so-called on-shell renormalisation scheme \cite{Schwartz:2014, Peskin:1995ev} the magnetic moment reads
\begin{equation}
	\vec{\mu}=\frac{Q_{f}}{ m_{f}}\biggr[1+F_2(p^2)\Big|_{p^2 = 0}\,\biggl]\, \frac{\vec{\sigma}}{2} \quad\quad \text{with} \quad\quad F_2(p^2)\Big|_{p^2 = 0}=\frac{g-2}{2}  \eqqcolon\; a_f~,
\end{equation}
where we defined the anomalous magnetic moment (AMM) $a_f$ of the fermion $f$.
The first and by far largest contribution to a lepton AMM dates back to 1948 \cite{PhysRev.73.416} and originates from the correction of the fermion-photon vertex at $1$-loop (the so-called Schwinger term) and amounts to $a^{\scriptscriptstyle{(1-\mathrm{loop})}}_f=\frac{\alpha}{2\pi}\simeq 0.00116$, where $\alpha$ is the fine-structure constant \footnote{The first accurate experimental determination of the AMM of the electron $a_{e}$ discovered by Kush and Foley from Columbia University (1948) gave $a_{e}^{\text {exp}}(\text{Columbia})=0.00119(5)$ \cite{KU, KU1, KU2}.}.

On the theoretical side, there are further contributions that do not come from Quantum Electrodynamics (QED) and contribute to $a^{\mathrm{SM}}_{\mu}$. In the case of the muon $\mu$, these further quantum effects cannot be neglected, and the AMM results schematically from \cite{ParticleDataGroup:2016lqr}
\begin{equation}
	a^{\mathrm{SM}}_{\mu}=a^{\mathrm{QED}}_{\mu}+\underbrace{ a^{\mathrm{HPV}}_{\mu}+a^{\mathrm{HLbL}}_{\mu}}_{a^{\mathrm{hadronic}}_{\mu}}+a^{\mathrm{electroweak}}_{\mu}~.
	\label{eq:SMAMM}
\end{equation}
\begin{figure}[t]
\begin{center}
\includegraphics[width=1\textwidth]{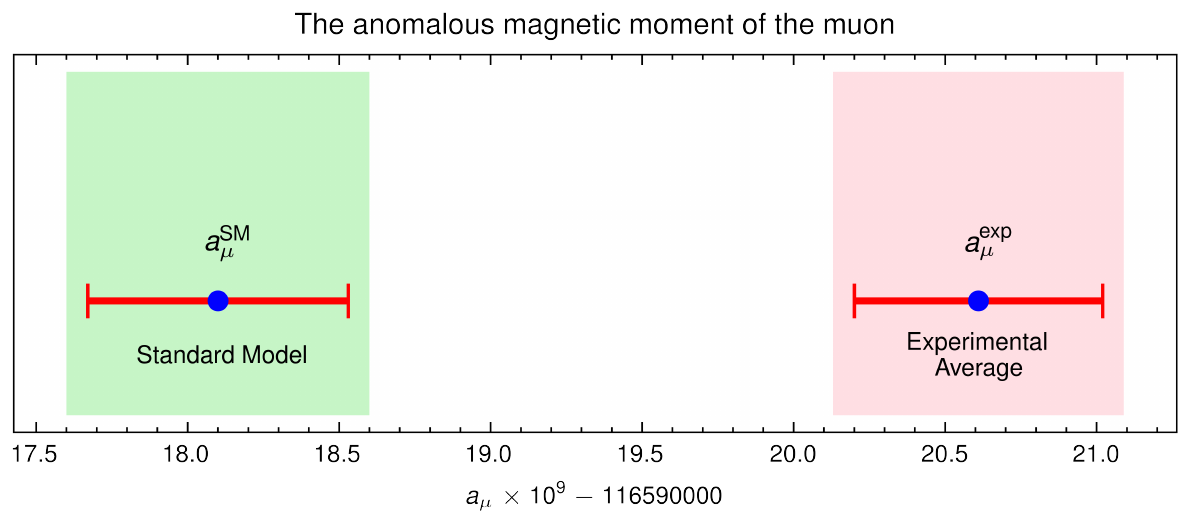}
\caption{ The disparity between the theoretically expected anomalous magnetic moment and the experimental average with the corresponding error bars. The gap between these points corresponds to the aforementioned $\sigma = 4.2$.}
\label{fig:AMM-muon}
\end{center}
\end{figure}
%
%
The AMM $a_{\mu}^{\mathrm{QED}}$ includes photon-fermion corrections calculated with an accuracy of up to $5$-loop level and amounts to $a_{\mu}^{\mathrm{QED}}= 116 584 718.931(104) \times 10^{-11}$ \cite{Kataev:2006, Passera:2004, Kurz:2016, Laporta:2017, Aoyama2017, Kinoshita:2004, Aoyama2012, Aoyama2020}. The hadronic part of (\ref{eq:SMAMM}) yields $a_{\mu}^{\mathrm{HPV}}=\SI{6845(40)e-11}{}$ \cite{HPV1,HPV2,HPV3,HPV4,HPV5,HPV6,HPV7}, which takes into account virtual strong-interacting effects with the leading contribution arising from hadronic vacuum polarisation (HVP), while $a_{\mu}^{\mathrm{HLbL}}=\SI{92(18)e-11}{}$ corresponds to hadronic light-by-light (HLbL) scattering contributions \cite{Melnikov:2003xd, Masjuan:2017tvw,Colangelo:2017fiz,Hoferichter:2018kwz,Gerardin:2019vio, Bijnens:2019ghy,Colangelo:2019uex,Pauk:2014rta,Danilkin:2016hnh,Knecht:2018sci,Eichmann:2019bqf, Roig:2019reh}. Weak processes involving $W$, $Z$ and Higgs bosons yield  $a_{\mu}^{\mathrm{weak}}=153.6(1.0)\times\SI{e-11}{}$ \cite{weak2, weak}. All together, this leads to a theoretical value \footnote{Expressed by the g-factor this results in $g^{\mathrm{SM}}_{\mu}=\SI{2.00233183620(86)}{}$.} \cite{Aoyama2020}
\begin{equation}
	 a^{\mathrm{SM}}_{\mu}=\SI{116 591 810(43)e-11}{}~.
	 \label{eq:AMMth}
\end{equation}
On the experimental side, however, there have been tremendous achievements in the high-precision measurement of the AMM of the muon. The latest interim results of the Fermilab National Accelerator Laboratory (FNAL) Muon $g-2$ Experiment (E$989$) for the muon magnetic anomaly determine $a^{\mathrm{exp}}_{\mu}(\mathrm{FNAL})=\SI{116 592 040(54)e-11}{}$ \cite{PhysRevLett.126.141801}. Combining these latest results with the previous experiment E$821$ from the Brookhaven National Laboratory (BNL), published in 2004, we find a new experimental average for the AMM of \footnote{The experimental result in this case was $a^{\mathrm{exp}}_{\mu}(\mathrm{BNL})=116592091(63) \times 10^{-11}\;\left[0.54\;\mathrm{ppm}\right].$}
\begin{equation}
	a^{\mathrm{exp}}_{\mu}=\SI{116592061(41)e-11}{}~,
	\label{eq:AMMexp}
\end{equation}
with a precision of $(0.35\; \mathrm{ppm})$\footnote{The interim result is analysed from the Run-$1$ dataset of $2018$. Later runs are being evaluated during the writing of this paper, and Run-$6$ is being planned. Furthermore, the E$24$ experiment at J-PARC will start in $2024$, which promises further accuracy \cite{Otani:2018kpl}.}\footnote{The experimental value for the AMM of the muon corresponds to $g_{\mu}^{\mathrm{exp}}=\SI{2.00233184122(82)}{}$}.
Comparing \eqref{eq:AMMth} and \eqref{eq:AMMexp} we find 
\begin{equation}
\d a_{\mu} =a_{\mu}^{\mathrm{exp}}-a_{\mu}^{\mathrm{SM}}= \SI{251(59)e-11}{}~,
\label{eq:da_DATA}
\end{equation}
which corresponds roughly to a discrepancy of $4.2$ standard deviations. The AMM for the SM and the experimental average are depicted in Fig.~\ref{fig:AMM-muon}. 

The origin of this discrepancy is still unknown and traditionally holds the possibility of \say{new physics}. In this paper, we want to shed some light on this yet uncharted path by explicitly examining the contribution $a^{\mathrm{electroweak}}_{\mu}$ in more detail, focusing on the Higgs sector. Moreover, we consider the possibility of more massive copies of the Higgs particle, offered by intersecting brane world scenarios in the context of type IIA superstring theory \cite{Kiritsis:2002aj, Anastasopoulos:2005ba, Armillis:2008bg,
Anchordoqui:2021llp, Anchordoqui:2021vrg, Anchordoqui:2021lmm,  AKKM}. 


\subsection{Intersecting D-branes and light stringy states}
The work carried out in this paper is based on semi-realistic \emph{intersecting D-brane configurations}. In this framework, the SM particles arise as fluctuations of a ``local set of D-branes'', where \textit{local} means a set of D-branes spanning the four-dimensional Minkowski space and wrapping various cycles of the internal compact six-dimensional manifold. They intersect in an area in transverse space, whose linear size is of the order of or smaller than the string length so that the SM is localised in a specific area in the internal space. We will assume this internal manifold to be a six-dimensional torus $\mathbb{T}^6$, which can be factorised into three $2$-tori $\mathbb{T}^2$: $\mathbb{T}^6=\mathbb{T}^2_1\times \mathbb{T}^2_2\times \mathbb{T}^2_3$.

In this context, the Standard Model particles correspond to the lightest fluctuations of open strings stretched between the stacks of D-branes~\cite{Inter_Dbrane_1,Inter_Dbrane_2,Inter_Dbrane_3,Inter_Dbrane_4,Inter_Dbrane_5,Inter_Dbrane_6,Inter_Dbrane_7,Inter_Dbrane_8,Inter_Dbrane_9,Inter_Dbrane_10,Inter_Dbrane_11,Anastasopoulos:2006da, Anastasopoulos:2012zu, Anastasopoulos:2015bxa,Inter_Dbrane_12}\footnote{And the review \cite{Kiritsis:2003mc}.}. This, as a result, realizes the SM spectrum in terms of adjoint and bi-fundamental representations of the local D-brane gauge groups\footnote{This is also the same requirement so that the SM spectrum is such that it can be coupled to any hidden sector in terms of bi-fundamental messengers. This is an important ingredient in models of emergent gravity \cite{Betzios:2020sro}, where axions \cite{Anastasopoulos:2018uyu}, graviphotons/dark-photons \cite{Betzios:2020sgd, Anastasopoulos:2020xgu}, neutrinos \cite{Anastasopoulos:2022sji} also emerge with special properties \cite{Anastasopoulos:2020gbu, Anastasopoulos:2021osp}.}. In particular, the general rules can be summarised as follows:
\bi
\item[(a)] Strings with both ends on a stack of $N$ parallel D-branes transform in the adjoint of $U(N) \simeq SU(N)\times U(1)_N$. They give rise to non-abelian $U(N)$ gauge bosons living on the worldvolume of this stack of branes.
\item[(b)] Strings stretched from a stack of $N$ to a stack of $M$ D-branes transform in the bi-fundamental representation $(\mathbf{N};\mathbf{\overline M})_{+1,-1}$ of $U(N)\times U(M)$, where the $\pm 1$ subscripts are the charges under the abelian parts of the gauge group. They give rise to four-dimensional chiral fermions living in the common worldvolume of the two stacks of branes.
\ei
The SM's gauge group is typically described by one stack of three branes, one stack of two branes, and a given number $k$ of single D-branes, giving rise to gauge groups $U(3)\times U(2)\times \prod_{i=1}^k U(1)_i$. The non-abelian parts of the three and two D-brane stacks give the $SU(3)\times SU(2)$, and the hypercharge $Y$ is a linear combination of the abelian factors living on each stack. Such embedding of the SM predicts several additional abelian factors which are ``superficially'' anomalous\footnote{The theory is anomaly-free in the UV but ``seems'' anomalous in the IR.}, and their anomalies are canceled by the \textit{Green-Schwarz mechanism} and generalised \textit{Chern-Simons} terms and become massive \cite{GS_1,GS_2,GS_3,
Anastasopoulos:2006cz,
Anastasopoulos:2007fgu,
Anastasopoulos:2008jt,
AKKM}. In these models, the masses and Yukawa couplings depend on various parameters related to the geometry of the internal manifold \cite{Anastasopoulos:2016yjs, AKT, AKRT, ANASTASOPOULOS2022115749}.
\begin{figure}[t]
        \centering
        \includegraphics[width=.4\textwidth]{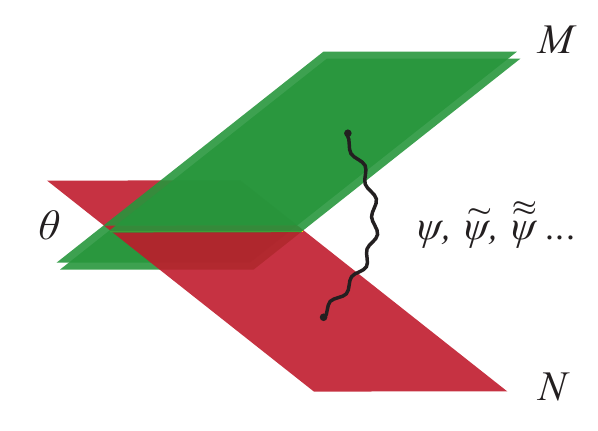}
    \caption{On each intersection lives a whole tower of states with the same quantum numbers. The lowest mode is typically massless and all the rest are massive, with masses $m^2_n=n\theta/\pi M^2_s$.}
    \label{fig:lightstringy}
\end{figure}

\begin{figure}[t]
        \centering
        \includegraphics[width=1\textwidth]{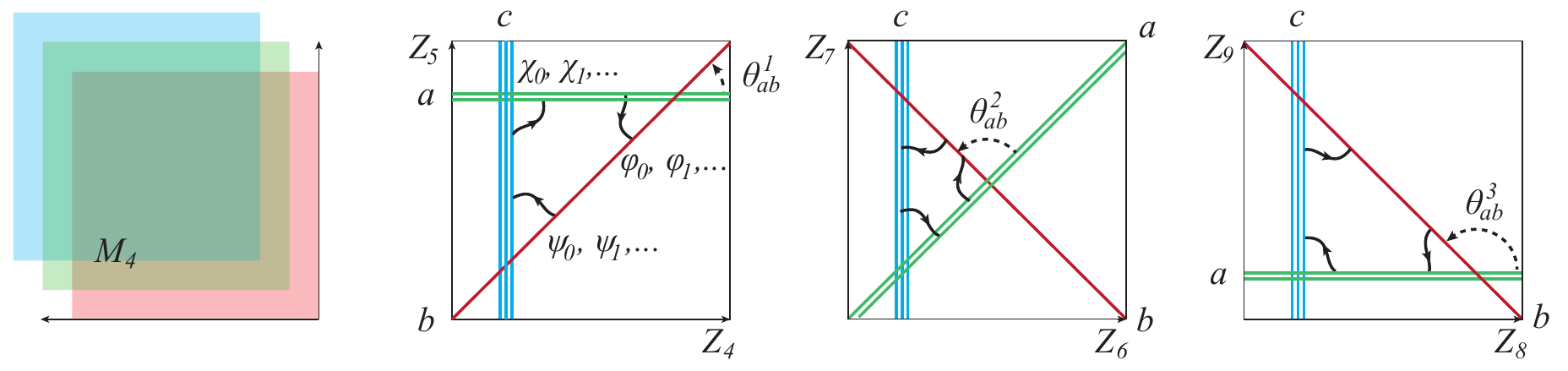}
    \caption{Three stacks of branes spanning 4-dimensional Minkowski space and intersecting in the internal manifold $\mathbb{T}^6=\mathbb{T}^2_1\times \mathbb{T}^2_2\times \mathbb{T}^2_3$. Each intersection has a massless zero mode, denoted with the subscript $0$, and an infinite tower of massive states whose first mode is denoted with the subscript $1$. At a given intersection, the zero modes are the same in each $2$-tori, but the towers are different for each $\mathbb{T}^2_i$.}
    \label{fig:3intersectingbranes}
\end{figure}

These stacks of D-branes intersect in the internal space with angles, which we will generically call $\theta\equiv\pi a$, with $a\in[0,1]$. Apart from the lowest (massless) modes –– which describe the SM fields –– there is at each intersection a whole tower of massive copies of the massless strings, with masses given by
\bea
m^2_n=n a M^2_s~,
\label{stringyVibrations}
\eea
where $n$ is an integer number and $M_s$ the string scale (Fig. \ref{fig:lightstringy}) \cite{Anastasopoulos:2015dqa,
Anastasopoulos:2016cmg, 
Anastasopoulos:2017tvo, 
Anastasopoulos:2018sqo}. This framework is schematically depicted in Fig.~\ref{fig:3intersectingbranes}, for the simplest configuration containing one stack of three coincident branes, one stack of two coincident branes and a single brane intersecting in the internal manifold. If the intersection angle $a$ is very small, then such massive copies have masses much lower than the string scale, and assuming further that the string scale is low, such \emph{light stringy states} become one of the first candidates to observe stringy phenomena in upcoming high energy experiments, e.g. the Forward Physics Facility (FPF) is planned to operate near the ATLAS interaction point during the LHC high-luminosity era \cite{FPF}. 

This paper aims to study the contributions of such light stringy states to the $g-2$ of the muon (we will denote it as $(g-2)_{\mu})$, focusing on the stringy excitations of the Higgs field. In D-brane semi-realistic vacua, the Higgs field is described by a string stretched between a stack of two and a single D-brane, and we will be interested in the excitation modes of such string. For this purpose, we will assume that:
\bi
\item The lightest excitations of all SM particles are the excitations of the Higgs particle, therefore living in the intersection with the smallest angle in this configuration.
\item Calling $ab$ the sector where this Higgs lives, we assume that the angle $\theta^1_{ab}$ is much smaller than the angles $\theta^2_{ab}$ and $\theta^3_{ab}$ (see Fig.~\ref{fig:3intersectingbranes}). The most significant contributions to the $(g-2)_{\mu}$, therefore, come from the tower of massive states living in the $ab$ intersection in the first $2$-torus $\mathbb{T}^2_1$.
\item These massive modes do not get a vacuum expectation value. Their stringy mass does not allow other minima in their potential. 

\item They live in supersymmetric configuration: the angles between the D-branes satisfy supersymmetric conditions, reminded for completeness in Eq.~\eqref{concrete setup 2} of the Appendix \ref{app:a}. However, we assume that supersymmetry is broken in some other sector, and it is mediated to this corner of the model allowing for a single real scalar and its copies to be the lightest stringy excitations.
\ei

\subsection{Results and outlook}

In the following we would like to briefly outline the results of our fruitful work.
The Yukawa couplings –– calculated in \cite{Anastasopoulos:2016yjs} –– can be used to evaluate the contributions of the excited Higgs states to the $(g-2)_{\mu}$ and explore the parameter space of the internal/fundamental theory.

The mass and the Yukawa couplings of the stringy excitations of the Higgs field depend on the structure of the internal space and the way the SM D-branes intersect.

For these excitations to significantly contribute to the $(g-2)_{\mu}$, their masses and couplings should be at a certain range in the parameter space, which sets some bounds on the size of the internal dimensions and the intersection angles. With this work, we show that there is a lower bound of order $\mathcal{O}(10^{-8}~{\rm GeV}^{-1})$ for the compact dimension where the branes intersect, and some intersection angles should be very small (of the order of $10^{-10}$). 
Here, we would like to make few comments regarding the small angle found and the viability of such configuration. 
\bi

\item 
From a String Theory point of view, the only requirements a semi-realistic D-brane configuration should satisfy are the tadpole cancellation conditions. These are global conditions which ensure that the total Ramond-Ramond charges must vanish in a compact space. In the framework of type IIA orientifold compactification with D6-branes and orientifold O6-planes, these conditions depend only on the number of branes in each stacks and on the different intersecting numbers of the D-branes, and is crucially independent of the value of the intersection angles of the branes \cite{Cvetic:2009yh}. The small value for some intersection angles that we get in this work is thus not hampered by any top-down theoretical constraints. The situation is analogous for the consistency condition which ensures that the hypercharge remains massless in four dimensions \cite{Cvetic:2009yh}, which can be implemented for any value of the intersection angles.

\item 

From a phenomenological point of view, 
we should first highlight that apart from the Higgs field and its tower of excitations, we can request that no other field lives in this intersection. All other matter fields correspond to open strings stretched between different D-brane stacks and they are unaffected by the value of this angle. Considering now the Higgs field and its excitations, the consequence of an ultra-small intersection angle is to shrink the mass gap between the different excitations. If the mass gap is tiny, it could lead to a breakdown of the effective field theory description of our model. 
However, for a string scale of the order of $10~{\rm TeV}$, the mass gaps from the first to the second and the second to the third excitations are of the order of hundreds of MeV ($103.5, 79.46,$ etc.). Therefore, stringy excitations of the Higgs are well separated even for small intersection angles, protecting from a breakdown of the effective field theory description of our D-brane model. 

\ei
Therefore, despite this ultra-small value, this proposal remains viable, both from a top-down and a bottom-up perspective.  


We also notice that there are areas in the parameter space of our model where higher excitations of the Higgs field give sizeable contributions to the $(g-2)_{\mu}$ and compete with the lower excitations. In these areas, however, the contribution of the Higgs excitations is much bigger than the accepted value for the $\d a_\m$ \eqref{eq:da_DATA}, and they are experimentally excluded.

In this work, the Yukawa terms of the excited Higgses are large enough in order to have sizeable contributions to the $(g-2)_{\mu}$. As a consequence, the lifetime of these light stringy excitations of the Higgs are very short. However, excitations of the same Higgs field living in other intersecting angles (from other tori, for example $a^2_{ab}$) might have very small Yukawas, which do not contribute to the $(g-2)_{\mu}$ but they have instead very extended lifetimes. In these cases, the same configuration presented here could also contain candidates for dark matter \cite{ANASTASOPOULOS2022115749}.

Finally, we explore semi-realistic D-brane configurations which can accommodate our setup. We focus on models where the Yukawa between the muon (and the tau) are present at tree level. However, the Yukawa coupling of the electron comes at higher order, by the presence of additional scalar fields or D-instantons \cite{Ibanez:2006da, Anastasopoulos:2009nk, Cvetic:2009yh, Cvetic:2009ez, Cvetic:2009ng, Camara:2010zm, Anastasopoulos:2010ca, Anastasopoulos:2010hu, Anastasopoulos:2011zz}. A second condition to be imposed is that the intersection where lives the Higgs which couples to the muon (and the tau) does not contain another SM fields, in order to avoid unrealistic light excitations for other SM particles. We find some configurations with a minimum number of four stacks of branes that have the above-mentioned pattern. In all of them, the discrepancy in the $(g-2)_{\mu}$ can be explained by the light-stringy excited Higgs states. Therefore, our phenomenological analysis can be realised in some semi-realistic D-brane configurations.

This paper is organized as follows. In Section \ref{Generic Scalar}, we provide the contribution of a single real scalar to the $(g-2)_{\mu}$. We use this contribution to relate this scalar's Yukawa coupling to the muon with its mass. In Section \ref{sec: excited Higgs}, we find areas of the parameter space where the excited Higgs can contribute significantly to the $(g-2)_{\mu}$.
In this section, we also evaluate the contribution from the double excited Higgs, and we use our analysis to set some further bounds to our parameter space.
In Section \ref{D-brane configurations}, we search within semi-realistic D-brane vacua studied in the past to find patterns that match the setup in the previous sections. Finally, Appendix \ref{app:a} summarizes the local supersymmetric conditions as well as the fermionic and bosonic modes living in the intersections, together with their vertex operators.

\section{A generic scalar contribution to the $(g-2)$ of the muon}\label{Generic Scalar}

In order to study the implication and inference of the stringy excitations of the Higgs field, we first consider the scalar contribution of a generic scalar field $\varphi$ to the $(g-2)_{\mu}$.

\begin{figure}[t]
        \centering
        \includegraphics[width=0.33\textwidth]{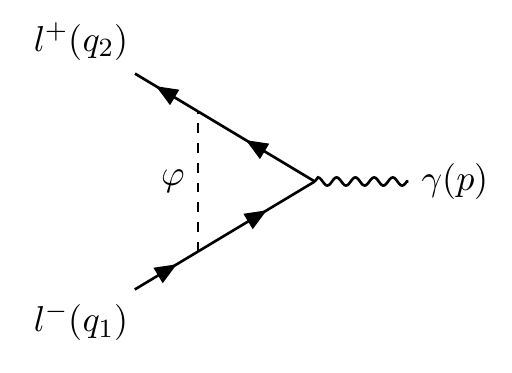}
    \caption{One-loop contribution from a real scalar $\varphi$ to the anomalous magnetic moment of a fermion $l$.}
    \label{fig:scalar_1_loop_contrib}
\end{figure}

For this purpose, let us consider the QFT described by the Lagrangian density
\begin{equation}
	\mathcal{L}=\mathcal{L}_{\mathrm{QED}}+\mathcal{L}_{\mathrm{Yukawa}}~,
	\label{eq:lagrangian}
\end{equation}
with 
\begin{equation}
	\mathcal{L}_{\text {Yukawa}}=\frac{1}{2}\left(\partial^{\mu} \varphi\right)\left(\partial_{\mu} \varphi\right)-\frac{m_{\varphi}^{2}}{2} \varphi^{2}-\lambda_{\ell}\varphi \bar{\ell} \ell~,
\end{equation}
where spinor QED for a given fermion $\ell$ of mass $m_{\ell}$ and charge $Q_{\ell}=1$ is supplemented by a real scalar field $\varphi$ of mass $m_\varphi$ coupled to the fermion via a Yukawa interaction with coupling constant $\lambda_{\ell}$. The AMM $a_{\ell}$ of the fermion $\ell$ can be extracted at all orders of perturbation theory from the scattering process $\ell^{+}\ell^{-}\rightarrow\gamma$. Using the Gordon identity, the amplitude $i\mathcal{F}^{\mu}$ for such process can be parametrized according to 
%
\begin{equation}
	i \mathcal{F}^{\mu}=-i e \bar{u}_{s_{2}}\left(q_{2}\right)\left[F_{1}\left(p^{2}\right) \gamma^{\mu}+F_{2}\left(p^{2}\right) \frac{i \sigma^{\mu \nu}}{2 m} p_{\nu}\right] u_{s_{1}}\left(q_{1}\right)~,
\end{equation}
where $\sigma^{\mu\nu}\equiv\frac{i}{2}[\gamma^{\mu},\gamma^{\nu}]$. The normalization is chosen such that at tree level the two \emph{Pauli form factors} $F_1$ and $F_2$ are $F_{1}\left(p^{2}\right)=1$ and $F_{2}\left(p^{2}\right)=0$. As described in Section~\ref{sect:introduction}, $F_{2}$ evaluated at $p^{2}=0$ gives the anomalous magnetic moment $a_{\ell}$ of the fermion $\ell$. This yields a UV-finite expression for $F_{2}\left(p^{2}\right)$. The one-loop contribution $\delta a_{\ell}$ to the AMM $a_{\ell}$ due to the exchange of a virtual scalar particle of mass $m_\varphi$ can then be extracted from the Feynman diagram depicted in Fig.~\ref{fig:scalar_1_loop_contrib}.

According to the corresponding Feynman rules, the final result reads
\bea
\d a_{\ell}={\l_{\ell}^2\over 8 \p^2}\left({m_{\ell}\over m_\varphi}\right)^2 \int_0^1 dz \; {(1 + z) (1 - z)\over
    (m_{\ell}/m_\varphi)^2 (1 - z)^2 + z} ~.
\label{da_GenericScalar}
\eea

Let us first set $\ell\equiv\mu$ as the muon and $\varphi\equiv h$ as the SM Higgs. Using $m_{\mu}=105~{\rm MeV}$, $m_{h}=125~{\rm GeV}$ as well as the Yukawa coupling between $h$ and $\mu$ \cite{Peskin:1995ev}, 
\bea\label{eq:SM_Higgs_yukawa}
\lambda_{\mu} \simeq 4\sqrt 2 \times 10^{-4},
\eea
we get from the integral \eqref{da_GenericScalar} the 1-loop contribution of the SM Higgs on the $(g-2)_{\mu}$:
\begin{equation}
	\delta a_{\mu}^{(h)}\simeq 3.9 \times 10^{-14}~,
\label{H data}
\end{equation}
which is negligible compared to the discrepancy \eqref{eq:da_DATA}. Any scalars providing a significant contribution to the $(g-2)_{\mu}$, if they exist, must have masses much lower than the SM Higgs. This can be obtained consistently with the current experimental bounds by considering scalars that couple only to the muon. In such cases, experimental bounds are weak, allowing scalars that can be as light as $2m_{\mu}$ \cite{Batell:2017kty,Davoudiasl:2018fbb}. In the following, we will consider such light scalars coupling only to the muon, with masses in the region $m_{\varphi}=\mathcal{O}(10^2~{\rm MeV})$.

\begin{figure}[t]
        \centering
        \includegraphics[width=1\textwidth]{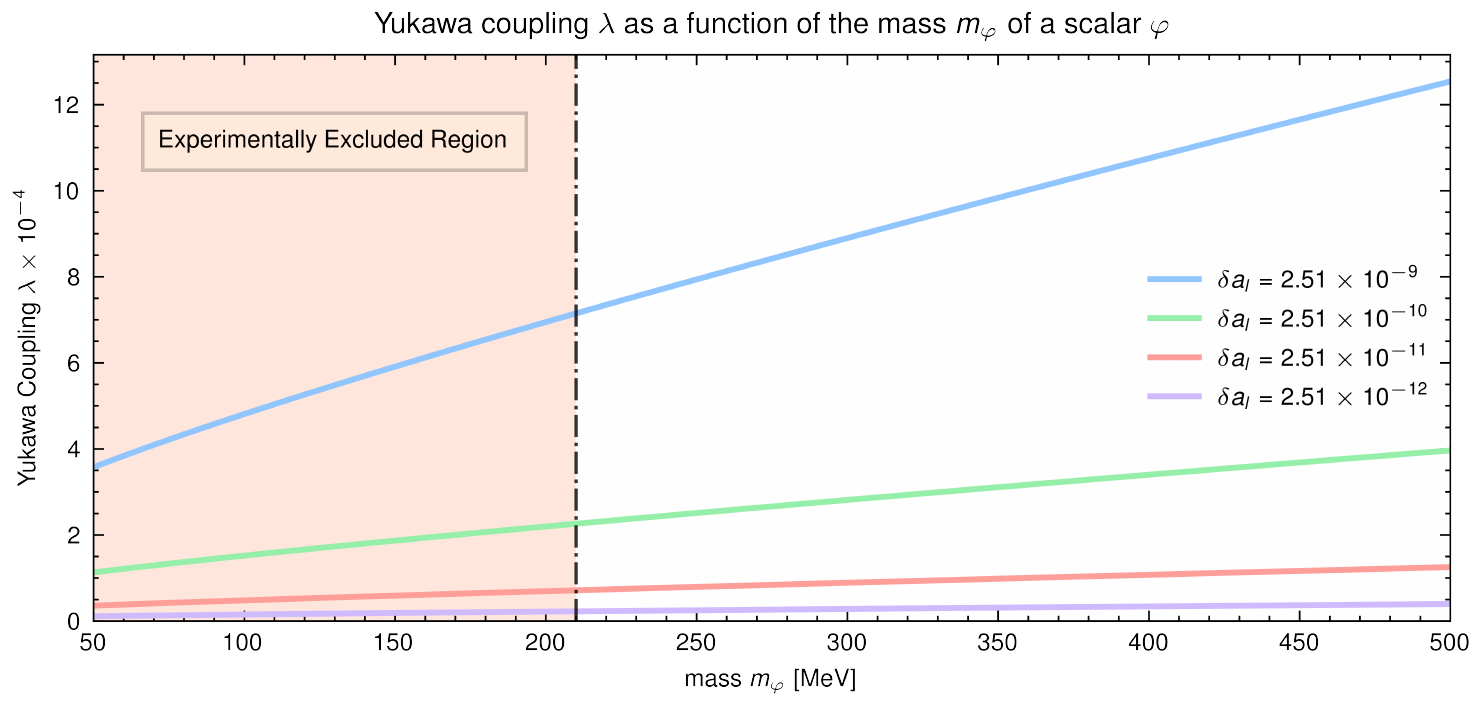}
    \caption{Plots of the Yukawa coupling $\l_{\m}$ between the muon and a given scalar $\varphi$, as a function of the scalar mass $m_\varphi$, in order for the 1-loop contribution to the $\d a_\m$ to be $2.5\times (10^{-9},10^{-10},10^{-11},10^{-12})$.}
    \label{fig:lvsM}
\end{figure}

From the integral \eqref{da_GenericScalar}, we can express the Yukawa coupling $\l_{\ell}$ in terms of the mass of the scalar $m_\varphi$, for a given 1-loop contribution $\d a_{\ell}$ to the AMM $a_{\ell}$ of $\ell$:
\bea
\l_{\ell}= 2\p \sqrt{2 \d a_{\ell}} {m_\varphi\over m_{\ell}} \left(\int_0^1 dz \;{(1 + z) (1 - z)\over
    (m_{\ell}/m_\varphi)^2 (1 - z)^2 + z} \right)^{-1/2}.
\label{eq:lvsM}
\eea
Setting $\ell\equiv\m$ for which $m_{\mu}=105~{\rm MeV}$, this function is plotted in Fig.~\ref{fig:lvsM} in terms of $m_\varphi$ varying in the $\mathcal{O}(10^2~{\rm MeV})$ region, for several contributions to the AMM of the muon $\d a_{\mu}=2.5\times (10^{-9},10^{-10},10^{-11},10^{-12})$. The round brackets indicate that the values contained therein are to be understood in the sense of a mathematical sequence. 

In such a range of masses for $m_{\varphi}$, we see that if $\varphi$ contributes to the total discrepancy $\d a_{\mu}=2.5\times 10^{-9}$, then the Yukawa coupling between the muon and $\varphi$ must be of order or bigger than the coupling between the muon and the SM Higgs $h$, which has been given in \eqref{eq:SM_Higgs_yukawa}. Since $\varphi$ will be associated with the excited states of the SM Higgs in the following, we require its Yukawa coupling with the muon to be smaller than \eqref{eq:SM_Higgs_yukawa}. We will therefore consider situations where such excited scalar modes give a significant contribution to $\delta a_{\mu}$, but which remains smaller than the total discrepancy $\d a_{\mu}=2.5\times 10^{-9}$. Typically, we will consider the case where one-tenth of the total discrepancy is absorbed by a scalar with a mass $m_{\varphi}=250~{\rm MeV}$, obtained for a coupling of order $2\times 10^{-4}$ according to Fig.~\ref{fig:lvsM}, which is indeed smaller than \eqref{eq:SM_Higgs_yukawa}. The aim of the next section is to study this situation in the framework of intersecting D-brane realization of the SM, where such scalar corresponds to one excitation of the open string whose lowest mode gives the Higgs field.

\section{Contribution of the excitations of the Higgs to the $(g-2)$ of the muon}
\label{sec: excited Higgs}

Let us consider the contribution to the $(g-2)_{\mu}$ coming from the first excited Higgs mode. Without loss of generality, we can assume that the smallest angle in this configuration is the $a_{ab}^1$ in the first torus $\mathbb{T}^2_1$, so that the lightest excitation of the Higgs, called $\tilde h$, has mass and coupling respectively given by \cite{Anastasopoulos:2016yjs}:
\begin{eqnarray}\label{eq:mass_1st_mode}
    m_{\tilde{h}}^2&=&a^1_{ab} M_s^2,\\
|\widetilde \l_{\mu} |&=&{|\l_{\mu} | \over \sqrt{\pi a^1_{ab}}} \Big( 2  \Gamma_{1-a^1_{ab},1-a^1_{bc},-a^1_{ca}} A^{(1)}_{h\mu {\mu}^C} M_s^2\Big)^{1/2},
\label{excitedHiggsCHARACTERISTICS}
\end{eqnarray}
where we have defined
\begin{equation}
	\Gamma_{a, b, c} \equiv \frac{\Gamma(a) \Gamma(b) \Gamma(c)}{\Gamma(1-a) \Gamma(1-b) \Gamma(1-c)}~,
\end{equation}
and $A^{(1)}_{h\mu {\mu}^C}$ is the area of the triangle formed by the three D-branes in the intersections of which live the Higgs and the left-handed and right-handed parts of the muon, in the first $2$-torus $\mathbb{T}^2_1$. 

\begin{figure}[t]
        \centering
        \includegraphics[width=140mm]{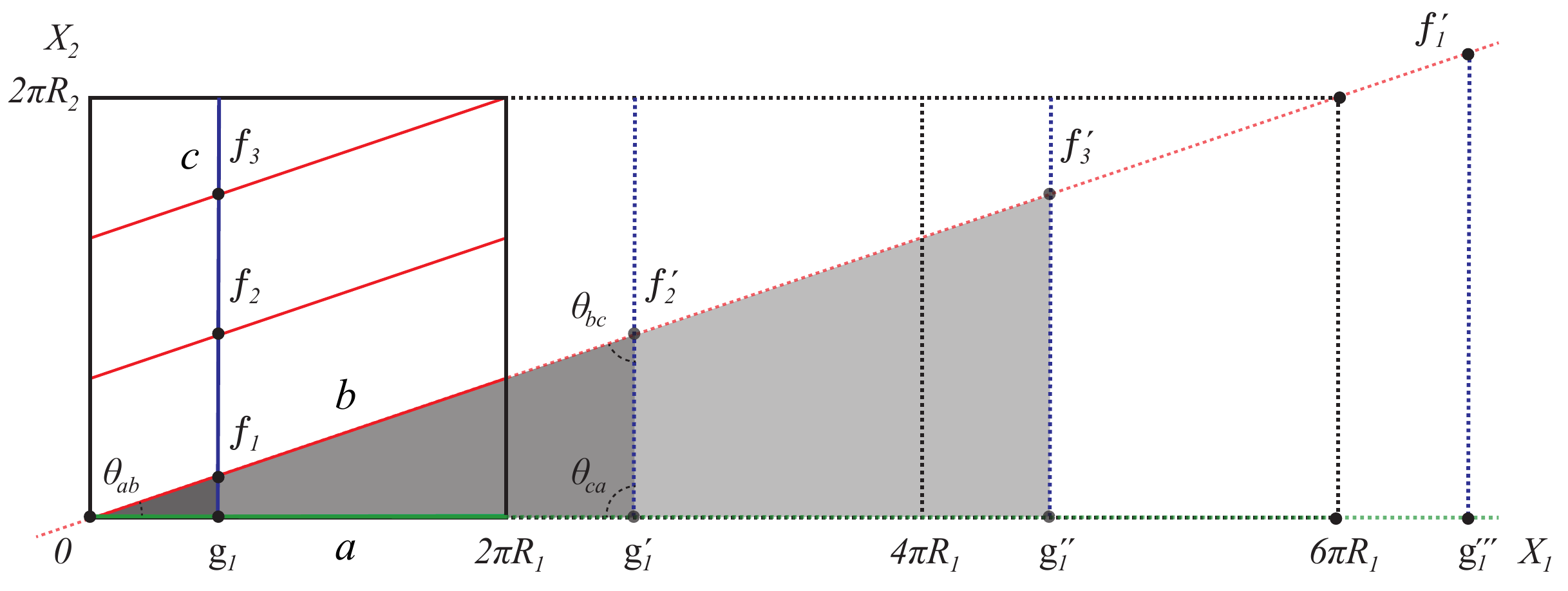}
    \caption{A general intersecting D-branes configuration with three stacks of branes $a$, $b$ and $c$ intersecting in the $i$\textsuperscript{th} $2$-torus $\mathbb{T}^2_i$ of the internal space $\mathbb{T}^6$. The Higgs is assumed to live in the $ab$ sector, and the left and right-handed fermions in the $bc$ and $ac$ sectors. We write the angles as $\theta_{ab}\equiv \pi a_{ab}$, with $a_{ab}\in[0,1]$, and similarly for the other intersections.}
    \label{fig:f}
\end{figure}

The generic expression for this area, in the case of three intersecting points $0, g^i_I$ and $f^i_J$ in the $i$\textsuperscript{th} $2$-torus $\mathbb{T}^2_i$, as depicted in Fig.~\ref{fig:f}, is given by \cite{Cvetic:2003ch,Anastasopoulos:2016yjs}
\begin{equation}
\label{eq:area}
  A_{\phi \chi^I\psi^J}^{(i)}(n)=\frac{1}{2}\left|\frac{\sin \pi a_{b c}^{i} \sin \pi a_{c a}^{i}}{ \sin \pi a_{a b}^{i}}\right| \left|f_{\chi^I \psi^J, i}(n)\right|^{2} ~,  
\end{equation}
with
\begin{equation}
    f_{\chi^I\psi^J,i}(n)=g^i_I -f^i_J + n_i \widetilde{L}^i_c ~.
\end{equation}
Here, the $g^i_I$ denote the points where the D-brane stacks $a$ and $c$ intersect in the respective $2$-torus $\mathbb{T}^{2}_i$. Analogously, the $f^i_J$ denote the intersection points in $\mathbb{T}^{2}_i$ for the D-brane stacks $b$ and $c$. The subindex $I$ denotes the number of possible intersections between the two D-branes. 
In the specific example of Fig.~\ref{fig:f} we have $I=1$ and $J=3$. Finally, $n_i$ denotes the wrapping number of the D-brane around the torus, and $\tilde L^i_c$ the length of the brane.
Since the Yukawa couplings are suppressed by terms of the form $e^{-A_{\phi \chi^I \psi^J}^{(i)}/(2 \pi \alpha^{\prime})}$, the dominant contribution to the Yukawas comes from the smallest areas, obtained for $I=J=1$ and $n_i=0$, as can be seen from Fig.~\ref{fig:f}. 
We will therefore restrict ourselves to this case, and denote $f_{\chi^1\psi^1,i}(n=0)\equiv f_{\chi\psi,i}$, where $\chi$ and $\psi$ will respectively denote the left-handed muon doublet and right-handed muon singlet in the following.

The result of the above analysis is that the area $A_{h\chi\psi}^{(i)}$ depends on the sines of various angles (this part cannot blow up since it involves angles from a triangle), and the length $f_{\chi\psi,i}$ depends on the size of the closed internal dimensions and can be considered of the order of the string scale up to a few micrometers\footnote{In this framework, the Yukawa coupling $\lambda_{\mu}$ between the muon and the zero mode of the scalar is given in terms of the geometric parameters by \cite{Cvetic:2003ch,Anastasopoulos:2016yjs}
\begin{equation}
\left|\l_{\mu}\right|= g_{\mathrm{op}}(2 \pi)^{-\frac{3}{4}}\left(\Gamma_{1-a_{a b}^{1}, 1-a_{b c}^{1},-a_{c a}^{1}} \Gamma_{1-a_{a b}^{2}, 1-a_{b c}^{2},-a_{c a}^{2}} \Gamma_{-a_{a b}^{3},-a_{b c}^{3},-a_{c a}^{3}}\right)^{\frac{1}{4}} \prod_{i=1}^{3} e^{-\frac{A_{\phi \chi \psi}^{(i)}}{2 \pi \alpha^{\prime}}}~,
\label{Yukawal}\end{equation}
where $g_{op}$ is the open string coupling. Since $\l_{\mu}$ depends on parameters (angles and distances on the D-branes) in all three tori, we can assume that its physical value given in \eqref{eq:SM_Higgs_yukawa} remains the same even if we modify some of the parameters in one torus. Parameters in the other tori can be adjusted so that $\l_{\mu}$ retains the value taken in \eqref{eq:SM_Higgs_yukawa}.}.

From the Yukawa term in \eqref{excitedHiggsCHARACTERISTICS} and the area \eqref{eq:area}, we can express the distance $f_{\chi\psi,1}$ on the D-brane as a function of the Yukawa couplings, the angles and the string scale as
\bea\label{eq:f}
|f_{\chi\psi,1}|=
\frac{\tilde \l_{\mu}}{\l_{\mu}}
\frac{1}{M_s}\Bigg(\frac{\pi a^1_{ab}}{\Gamma_{1-a_{ab}^1,1-a^1_{bc},a_{ab}^1+a^1_{bc}}}\left|\frac{\sin\pi a_{ab}^1}{\sin\pi a^1_{bc}\sin\pi(a_{ab}^1+a^1_{bc})}\right|\Bigg)^{1/2}.
\eea
In this expression, we have used the local supersymmetric condition, Eq.~\eqref{concrete setup 2}, in order to express the angle $a^1_{ca}$ in terms of $a^1_{ab}$ and $a^1_{bc}$. From $a^1_{ab}={m^2_{\tilde h}}/{M^2_s}$ we have an expression  for the distance which depends on the Yukawa couplings $\l_{\mu},\tilde \l_{\mu}$, the mass $m_{\tilde h}$, the string scale $M_s$ and the angle $a^1_{bc}$.

\begin{figure}[t]
        \centering
        \includegraphics[width=1\textwidth]{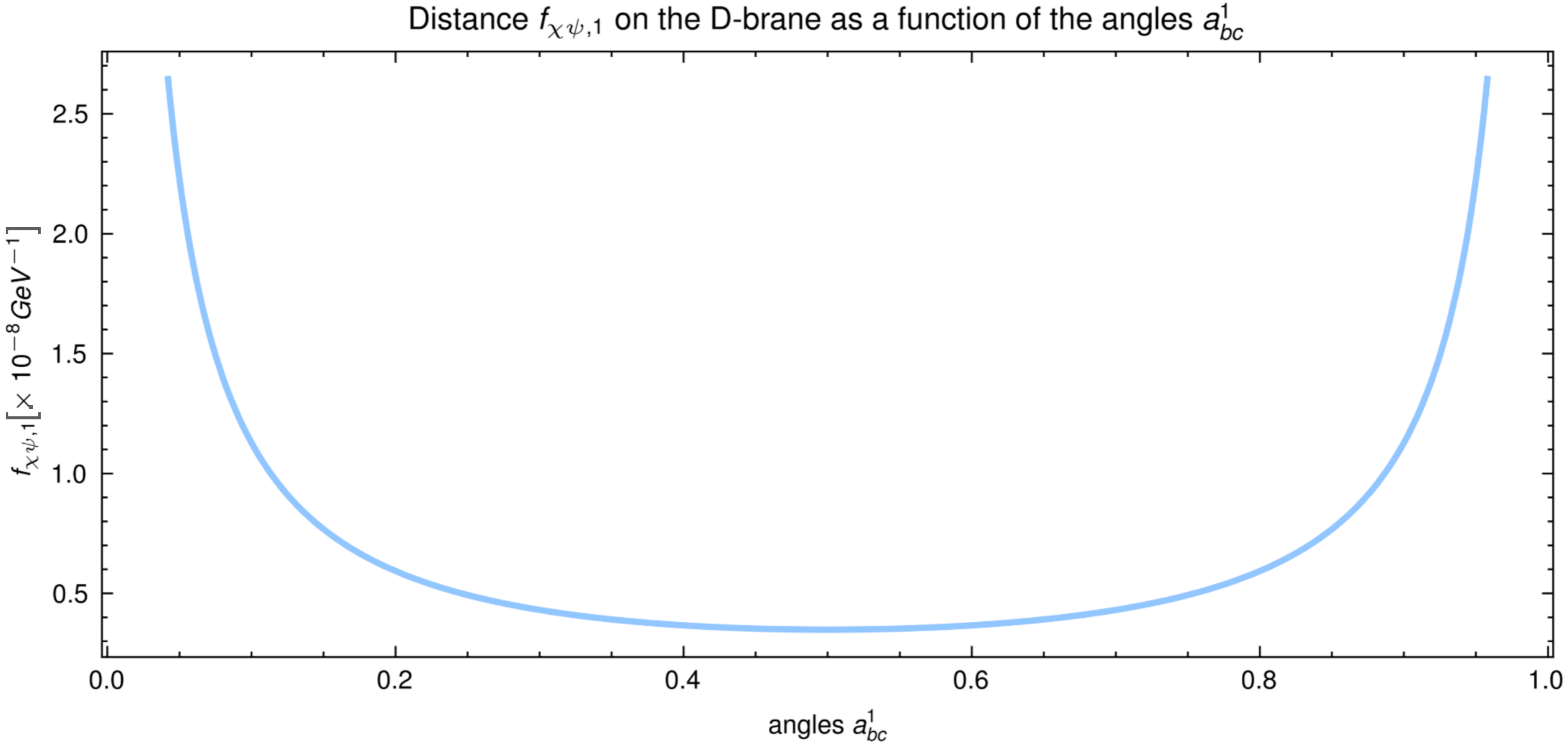}
    \caption{Length $f_{\chi\psi,1}$ in terms of the angle $a^1_{bc}$, for an angle $a^1_{ab}= 6\times 10^{-10}$ and a string scale $M_s=10~{\rm TeV}$, so that the excited Higgs $\tilde h$ has a mass $m_{\tilde h}=250~{\rm MeV}$ and contributes an amount of $2.5\times 10^{-10}$ to the $(g-2)_{\mu}$.}
    \label{fig:f(a)}
\end{figure}

As described in section~\ref{Generic Scalar}, a scalar field that couples only to the muon can have a mass as small as $2m_{\mu}$~\cite{Batell:2017kty,Davoudiasl:2018fbb}. In order to fix the ideas, we will consider the mass of the excited Higgs $\tilde h$ to be $m_{\tilde h}=250~{\rm MeV}$. For a string scale $M_s\simgt 10~{\rm TeV}$ \cite{Kiritsis:2003mc}, the relation \eqref{eq:mass_1st_mode} therefore sets an upper bound for the angle $a_{ab}^1$\footnote{
There are two different sources for the mass of the excited Higgs. One is due to the vibrations of the string \eqref{stringyVibrations}, and one is due to a coupling with the SM Higgs, proportinal to $g'v$. In this work, we assume that the second contribution is smaller than the stringy effects.
Let us note that turning on $g'$ for the mass of the excited Higgs in \eqref{excitedHiggsCHARACTERISTICS} will lead, for a fixed mass $m_{\tilde h}$, to a smaller value of $a^1_{ab}$, so that the bound \eqref{eq:angle_bound} is valid even when $g'\neq 0$.}
\bea\label{eq:angle_bound}
a^1_{ab}\simlt 6\times 10^{-10}.
\label{angle1}
\eea
According to the analysis following Fig.~\ref{fig:lvsM}, we request one-tenth of the full discrepancy \eqref{eq:da_DATA} to be absorbed by the excited Higgs $\tilde h$, namely $\d a_\m=2.5 \times 10^{-10}$. From the relation \eqref{eq:lvsM}, we then get a Yukawa coupling with the value:  
\bea
\tilde \l_{\mu}=2.5 \times 10^{-4}.
\label{tildel1}
\eea
Using the upper bound of \eqref{angle1} as well as \eqref{tildel1} and \eqref{eq:SM_Higgs_yukawa} in \eqref{eq:f}, we get the length $f_{\chi\psi,1}$ in terms of the angle $a_{bc}^1$ and plot the resulting function $f_{\chi\psi,1}(a_{bc}^1)$ in Fig.~\ref{fig:f(a)}. For any $a_{bc}^1\in[0,1]$, one gets a length $f_{\chi\psi,1}$ of order $10^{-8}~{\rm GeV}^{-1}\sim 10^{-21}~{\rm mm}$. Let us note that the value of $f_{\chi\psi,1}$ simply gives a lower bound for one size (the radius $R_2$ in the notations of Fig.~\ref{fig:f}) of the extra dimension of the first $2$-torus $\mathbb{T}^2_1$ where this excited Higgs $\tilde h$ lives: $R_2$ must be larger than $f_{\chi\psi,1}$, but then can be as large as the ${\rm TeV}^{-1}$ without any additional constraints coming from this analysis.

\begin{figure}[t]
        \centering
        \includegraphics[width=1\textwidth]{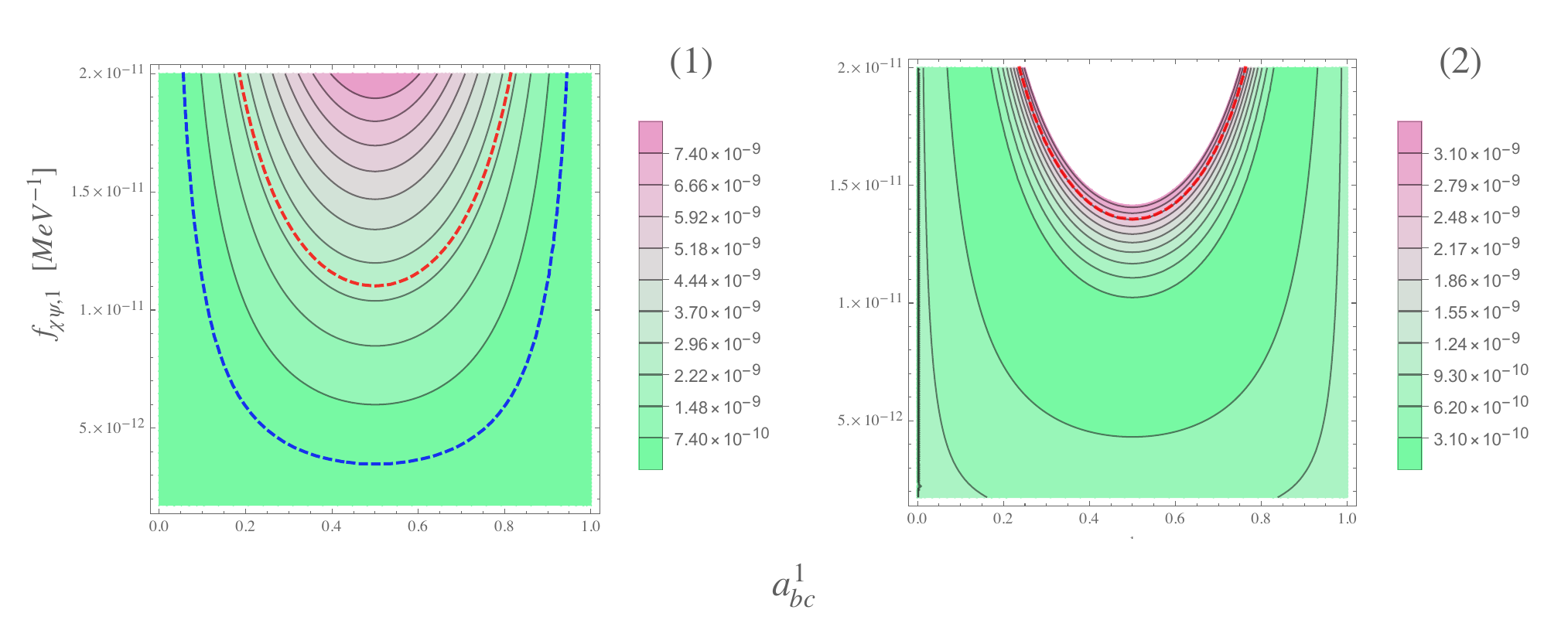}

    \caption{(1) The $\d a_\m$ contribution coming from an excited Higgs living in the intersection $ab$ with an angle $a^1_{ab}=6\times 10^{-10}$ and a string scale at $M_s=10~{\rm TeV}$. The blue line shows the plot of Fig.~\ref{fig:f(a)} embedded in this result, and the red line the full discrepancy \eqref{eq:da_DATA}. (2) The contribution to the $\d a_\m$ coming from the double excited Higgs, living in the same intersection $ab$. Again the red line shows the full discrepancy \eqref{eq:da_DATA}.}
    \label{fig:daForfandaRelaxed}
\end{figure}

The plot of Fig.~\ref{fig:f(a)} gives us the range where the distance $f_{\chi\psi,1}$ is such that the excited Higgs $\tilde h$ absorbs one-tenth of $\delta a_{\mu}$, for a fixed angle $a^1_{ab}= 6\times 10^{-10}$ and a fixed string scale $M_s=10~{\rm TeV}$. Keeping these two parameters $a^1_{ab}$, $M_s$ fixed, one can then explore the full region $a^1_{bc}\in [0,1]$ and $f_{\chi\psi,1}\in [0.2-2]\times 10^{-11}~{\rm MeV}^{-1}$, which amounts to change the Yukawa coupling $\tilde \lambda_{\mu}$, and see what are the corresponding contributions to $\delta a_{\mu}$ in this region of the parameter space. The results are presented in the left panel of Fig.~\ref{fig:daForfandaRelaxed}. The blue line shows the plot of Fig.~\ref{fig:f(a)} embedded in the results, and the red line the full $(g-2)_{\mu}$ discrepancy \eqref{eq:da_DATA}. This plot in the left panel of Fig.~\ref{fig:daForfandaRelaxed} shows two different regions. First, where the angle $a^1_{bc}$ is close to $0$ and $1$, in which case the contribution from the excited Higgs $\tilde h$ to $\delta a_{\mu}$ is almost independent of the length $f_{\chi\psi,1}$, and remains of order $10^{-10}$. Next, where the angle $a^1_{bc}\sim 0.4-0.6$ (in this case the angle between the D-branes $bc$ is almost orthogonal): here the contribution of $\tilde h$ to $\delta a_{\mu}$ increases quickly with $f_{\chi\psi,1}$. This sets an upper bound for $f_{\chi\psi,1}$ around $f_{\chi\psi,1}\sim 1.1\times 10^{-11}~{\rm MeV}^{-1}$, where the full discrepancy \eqref{eq:da_DATA} is bridged.



Next, we use the same values of the angles, the string scale and the distance $f_{\chi\psi,1}$ to predict the mass and the Yukawa coupling of the next, double excited Higgs field $\tilde {\tilde h}$ in this region of the parameter space. In terms of the geometric parameters, they are respectively given by \cite{Anastasopoulos:2016yjs}:
\begin{eqnarray}
{m}_{\tilde {\tilde h}}^2&=&2 a^1_{ab} M_s^2,\\
|\tilde {\tilde \l}_{\mu} |&=&{\sqrt{2} \over \pi a^1_{ab}} ~|\l_\mu|~  \Gamma_{1-a^1_{ab},1-a^1_{bc},-a^1_{ca}} \Big(A^{(1)}_{\varphi\bar{\psi}\psi} M_s^2-\p/2\Big).
\end{eqnarray}
For the same D-brane configuration with $a^1_{ab}=6\times 10^{-10}$, $M_s=10~{\rm TeV}$, we evaluate the contribution of the double excited Higgs $\tilde {\tilde h}$ to the $\d a_\m$ in the region $f_{\chi\psi,1}\in [0.2-2]\times 10^{-11}~{\rm MeV}^{-1}$ and present our results in the right panel of Fig.~\ref{fig:daForfandaRelaxed}. The red line shows the full discrepancy \eqref{eq:da_DATA}, above which the region is phenomenologically excluded.

\begin{figure}[t]
        \centering
        \includegraphics[width=1\textwidth]{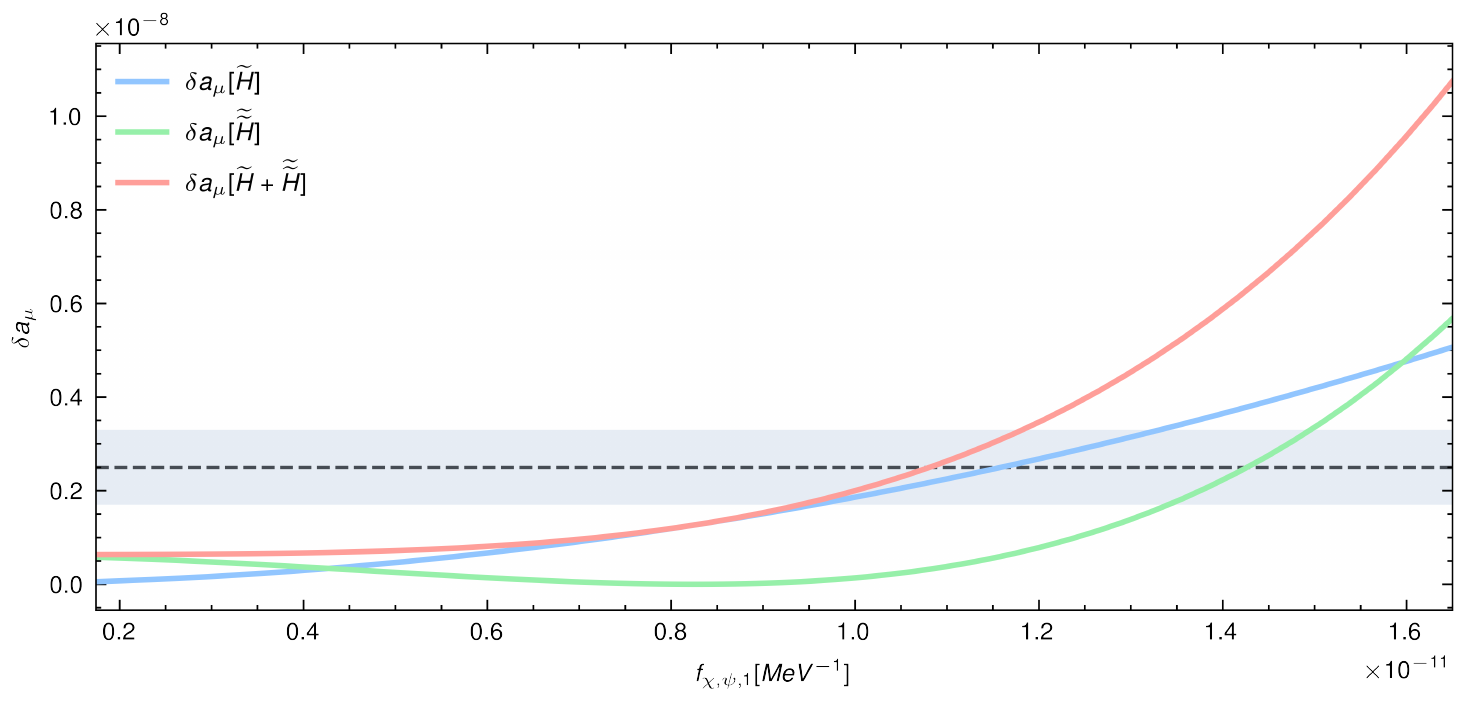}
    \caption{Contributions to the $\delta a_{\mu}$ coming from the first and second excited states $\tilde h$ and $\tilde{\tilde h}$, in terms of the length $f_{\chi\psi,1}$, for fixed angles $a^1_{ab}=6\times 10^{-10}$, $a^1_{bc}=0.5$ and string scale $M_s=10~{\rm TeV}$.}
    \label{fig:tildeHvstildetildeH}
\end{figure}

We should mention here that there are areas in the parameter space where the contribution of the double excited Higgs $\tilde{\tilde h}$ to the $(g-2)_{\mu}$ is larger than the contribution of the excited Higgs $\tilde h$. This basically comes from the fact that in these regions, the Yukawa coupling for $\tilde{\tilde h}$ is bigger than the Yukawa coupling for $\tilde h$, $\tilde {\tilde \l}_\mu> {\tilde \l}_\mu$. In order to present this effect, we choose a fixed value for $a^1_{bc}=0.5$, and plot in terms of $f_{\chi\psi,1}$ the contributions to the $\d a_\m$ coming from $\tilde h$ and $\tilde {\tilde h}$ separately, as well as their sum, together in Fig.~\ref{fig:tildeHvstildetildeH}. The blue line shows the $\tilde h$ contribution, the green one the $\tilde{\tilde h}$ contribution, the red one the sum of both, and the horizontal blue strip the $(g-2)_{\mu}$ discrepancy \eqref{eq:da_DATA} with its uncertainty.

From this plot, we see that the contribution coming from $\tilde{\tilde h}$ is bigger than the one coming from $\tilde h$ in two distinct regions. The first one is when $f_{\chi\psi,1}\simgt 1.6\times 10^{-11}~{\rm MeV}^{-1}$, which is phenomenologically excluded since in this region the different contributions are larger than the discrepancy \eqref{eq:da_DATA}. The second region is for a length $f_{\chi\psi,1}\simlt 4.25\times 10^{-12}~{\rm MeV}^{-1}$. In this case, the total contribution from $\tilde{\tilde h}$ and $\tilde h$ is smaller than the discrepancy \eqref{eq:da_DATA}, and this region of the parameter space cannot be excluded for now. A careful study of this region would require the computation of the Yukawa couplings for the higher excited modes of the Higgs, in order to see whether the total contribution converges or not\footnote{A divergent total contribution would signal a breakdown of the effective field theory description.}, and whether it remains smaller or of order of the full discrepancy \eqref{eq:da_DATA}, an analysis which is beyond the scope of this paper. The point we want to highlight here is that for the above considered parameters, the angles $a^1_{ab}=6\times 10^{-10}$, $a^1_{bc}=0.5$ and a string scale $M_s=10~{\rm TeV}$, there is a region $4.25\times 10^{-12}~{\rm MeV}^{-1}\simlt f_{\chi\psi,1} \simlt 1.1\times 10^{-11}~{\rm MeV}^{-1}$ where the contribution of $\tilde{\tilde h}$ is smaller than the one of $\tilde h$, such that their sum gives a significant contribution to the full discrepancy \eqref{eq:da_DATA}.

\section{Excited Higgs and semi-realistic D-brane configurations}
\label{D-brane configurations}

In this paper, we have evaluated the contribution of light stringy states to the $(g-2)_{\mu}$, focusing on the stringy excitations of a given scalar field localised at the intersection of two intersecting D-branes. In case such scalar couples only to the muon, it evades the more stringent experimental bounds we have in case of a coupling with the electron and the quarks. Experimental bounds being weaker, light excitation masses are allowed, and we have shown in this case that there exist regions of the parameter space where these light scalar excitations can provide significant contributions to the $(g-2)_{\mu}$.

This section aims to analyse and classify the local D-brane configurations that can realise this proposal. The crucial point is that these constructions must contain at least one Higgs doublet which only couples to the muon (and the tau) at tree level but not to the electron or the quarks. This assumption, therefore, requires tree-level Yukawa couplings between the muon and the Higgs field (and consequently to all massive copies), and no Yukawa between this Higgs and the electron and quarks. In this case, other scalars or instanton effects will provide the missing Yukawas of the electron and quarks, suppressing their values and giving the observed hierarchy of the lepton and quark masses
\cite{Ibanez:2006da, Anastasopoulos:2009nk, Cvetic:2009yh, Cvetic:2009ez, Cvetic:2009ng,
Anastasopoulos:2009mr,
Camara:2010zm,
Anastasopoulos:2010ca, Anastasopoulos:2010hu, Anastasopoulos:2011zz}. Denoting by $H_d$ the Higgs doublet which couples to the muon (and the tau), the D-brane constructions presented in this section must therefore satisfy the following pattern regarding the tree-level Yukawa couplings:
\begin{subequations}\label{eq:Yukawa_conditions}
\begin{align}
L_{\mu/\tau}l_{\mu/\tau}^CH_d~&:~\text{allowed}\\
L_{e}l_{e}^CH_d~&:~\text{not allowed}\\
Qd^CH_d~&:~\text{not allowed}.
\end{align}
\end{subequations}
In string theory language, the selection rules for the Yukawa coupling are encoded in the overall charge under the abelian factors living on the D-branes. 

A second condition to be imposed on our semi-realistic D-brane constructions is that no other SM field (which could only be the lepton doublet $L$) can live in the same intersection as the Higgs $H_d$ which couples to the muon (and the tau). If this would be the case, the ultra-small value for the intersection angle where $H_d$ lives would lead to light excitations for the leptons $L$, which is phenomenologically excluded. We thus impose that:
\begin{equation}\label{intersection_condition}
    \text{\emph{The lepton doublet $L$ should not share the same intersection with the Higgs $H_d$.}}
\end{equation}
In the following, we analyse different realizations of the SM, classified by the different ways that the hypercharge is described in such models, that satisfy both conditions \eqref{eq:Yukawa_conditions} and \eqref{intersection_condition}, based on the general classification and analysis presented in \cite{Anastasopoulos:2006da}.

\subsection{Three stacks models}
\label{sect:3_stacks}
The minimal intersecting D-brane configurations in which can be embedded the SM spectrum require three stacks of branes, giving rise to a gauge group $U(3)_a\times U(2)_b\times U(1)_c$. There are then two distinct hypercharge embeddings able to give the correct hypercharge to the SM particles:
\begin{equation}
    Q_Y=-\frac{1}{3}Q_a+\frac{1}{2}Q_b, \quad Q_Y=\frac{1}{6}Q_a+\frac{1}{2}Q_c.
\end{equation}
For the first case $Q_Y=-\frac{1}{3}Q_a+\frac{1}{2}Q_b$, the SM spectrum reads:
\begin{subequations}
\begin{align}
&Q& &(1,1,0)_{1/6}&\\
&u^C& &(2,0,0)_{-2/3}&\\
&d^C& &(-1,0,\epsilon_d)_{1/3}&\\
&L& &(0,-1,\epsilon_L)_{-1/2}&\\
&l^C& &(0,2,0)_{1}&\\
&H_d& &(0,-1,\epsilon_H)_{1/2}&\\
&H_u& &(0,1,\epsilon_{H})_{-1/2},&
\end{align}
\label{3.stack.models}
\end{subequations}
while for the second hypercharge embedding $Q_Y=\frac{1}{6}Q_a+\frac{1}{2}Q_c$, it is given by:
\begin{subequations}
\begin{align}
&Q& &(1,\epsilon_Q,0)_{1/6}&\\
&u^C& &(-1,0,-1)_{-2/3}&\\
&d^C& &(2,0,0)_{1/3}& &\text{or}& &(-1,0,1)_{1/3}&\\
&L& &(0,\epsilon_L,-1)_{-1/2}&\\
&l^C& &(0,0,2)_{1}&\\
&H_d& &(0,\epsilon_H,-1)_{-1/2}&\\
&H_u& &(0,\epsilon_{H'},1)_{1/2},&
\end{align}
\end{subequations}
with the hypercharge of each species indicated as a subscript. While the conditions \eqref{eq:Yukawa_conditions} can be satisfied by a suitable choice of the different $\epsilon$ parameters, one sees that the condition \eqref{intersection_condition} is violated in both models. There is thus no three stacks model which can phenomenologically accommodate our proposal of an ultra-small intersection angle explaining (part of) the $(g-2)_{\mu}$ discrepancy.

\subsection{Four stacks models}
\label{sect:4_stacks}
Considering four stacks models giving rise to a gauge group $U(3)_a\times U(2)_b\times U(1)_c\times U(1)_d$, there are $8$ different hypercharge embeddings able to reproduce the correct hypercharge for the SM particles, which have been classified in \cite{Anastasopoulos:2006da}. One famous choice, which gives the most important number of anomaly-free models, is given by
\begin{equation}\label{eq:Madrid_model}
    Q_Y=\frac{1}{6}Q_a+\frac{1}{2}Q_c+\frac{1}{2}Q_d,
\end{equation}
for which the SM spectrum reads:
\begin{subequations}
\begin{align}
	&Q& &(1,\epsilon_Q,0,0)_{1/6}&
	 & & & \\
	 \label{eq:uc_charges}
	&u^C& &(-1,0,-1,0)_{-2/3}&	&\text{or}&  &(-1,0,0,-1)_{-2/3}&\\
	&d^C& &(2,0,0,0)_{1/3}& &\text{or}& &(-1,0,1,0)_{1/3}& &\text{or}& &(-1,0,0,1)_{1/3}&\\
	&L& &(0,\epsilon_L,-1,0)_{-1/2}& &\text{or}& &(0,\epsilon_L,0,-1)_{-1/2}&\\
	&l^C& &(0,0,2,0)_{1}& &\text{or}& &(0,0,1,1)_{1}& &\text{or}& &(0,0,0,2)_{1}&\\
	\label{eq:Hu_charges}
	&H_u& &(0,\epsilon_H,0,1)_{1/2}& &\text{or}& &(0,\epsilon_H,1,0)_{1/2}&\\
	\label{eq:Hd_charges}
    &H_d& &(0,\epsilon_{H},0,-1)_{-1/2}& &\text{or}& &(0,\epsilon_{H},-1,0)_{-1/2}.&
\end{align}
\end{subequations}
As explained above, we need to choose the different quantum numbers of the matter species in such a way that tree-level Yukawa coupling for the muon $L_{\mu}l^C_{\mu}H_d$ is allowed, while the ones for the electron $L_{e}l^C_{e}H_d$ and the down quarks $Qd^CH_d$ are forbidden. There are two distinct possible choices for the charge assignments of $H_d$ as listed in \eqref{eq:Hd_charges}. Considering the first possibility $H_d(0,\epsilon_{H},0,-1)_{-1/2}$ leads to one possible model compatible with the above mentioned conditions \eqref{eq:Yukawa_conditions} and \eqref{intersection_condition}:
\begin{subequations}\label{eq:4_stacks_model_1}
\begin{align}
	&Q& &(1,\epsilon_Q,0,0)_{1/6}&	 & & & \\
	&d^C& &(2,0,0,0)_{1/3}& &\text{or}& &(-1,0,1,0)_{1/3}& &\text{or}& &(-1,0,0,1)_{1/3}&\\
	&L_e& &(0,\epsilon_H,-1,0)_{-1/2}&\\
    &L_{\mu}& &(0,-\epsilon_H,-1,0)_{-1/2}&\\
	&l^C_{e,\mu}& &(0,0,1,1)_{1}&\\
    &H_d& &(0,\epsilon_{H},0,-1)_{-1/2}.&
\end{align}
\end{subequations}
In this model, the Yukawas for the down quarks $Qd^CH_d$ are forbidden in the configurations where $d^C~(2,0,0,0)_{1/3}$ and $d^C~(-1,0,1,0)_{1/3}$ independently of the value of $\epsilon_Q$, while for $d^C~(-1,0,0,1)_{1/3}$ we must have $\epsilon_H=\epsilon_Q$ to forbid the coupling $Qd^CH_d$. 

For the second possibility $H_d(0,\epsilon_{H},-1,0)_{-1/2}$, one gets another model compatible with the conditions \eqref{eq:Yukawa_conditions} and \eqref{intersection_condition}:
\begin{subequations}\label{eq:4_stacks_model_2}
\begin{align}
	&Q& &(1,\epsilon_Q,0,0)_{1/6}&	 & & & \\
	&d^C& &(2,0,0,0)_{1/3}& &\text{or}& &(-1,0,1,0)_{1/3}& &\text{or}& &(-1,0,0,1)_{1/3}&\\
	&L_e& &(0,\epsilon_H,0,-1)_{-1/2}&\\
    &L_{\mu}& &(0,-\epsilon_H,0,-1)_{-1/2}&\\
	&l^C_{e,\mu}& &(0,0,1,1)_{1}&\\
    &H_d& &(0,\epsilon_{H},-1,0)_{-1/2}.&
\end{align}
\end{subequations}
In this model, it is for the configuration $d^C~(-1,0,1,0)_{1/3}$ that $\epsilon_H=\epsilon_Q$ must be imposed in order to forbid the Yukawa coupling $Qd^CH_d$, while for the two others $d^C~(2,0,0,0)_{1/3}$ and $d^C~(-1,0,0,1)_{1/3}$, $\epsilon_Q$ can remain arbitrary.

Tree level Yukawa coupling for the up quarks $Qu^CH_u$ can also be introduced. For the hypercharge embedding \eqref{eq:Madrid_model} we are considering here, there are two possible charge assignments for the up anti-quark singlets $u^C$ and Higgs doublets $H_u$, respectively given in \eqref{eq:uc_charges} and \eqref{eq:Hu_charges}. It is easy to see that only two combinations of them can allow a Yukawa coupling $Qu^CH_u$,
\begin{subequations}\label{eq:config_up_yukawa}
\begin{align}
	&Q& &(1,\epsilon_Q,0,0)_{1/6}&  && &Q& &(1,\epsilon_Q,0,0)_{1/6}& \\
	&u^C& &(-1,0,0,-1)_{1/3}& &\text{or}& &u^C& &(-1,0,-1,0)_{1/3}& \\
    &H_u& &(0,\epsilon_{H},0,1)_{1/2}& && &H_u& &(0,\epsilon_{H},1,0)_{1/2},&
\end{align}
\end{subequations}
provided that $\epsilon_Q=-\epsilon_H$. They can be implemented independently in the two different models listed above, except for the configuration where $d^C~(-1,0,0,1)_{1/3}$ in the model \eqref{eq:4_stacks_model_1} and $d^C~(-1,0,1,0)_{1/3}$ in the model \eqref{eq:4_stacks_model_2}, for which it has been required that $\epsilon_Q=\epsilon_H$. On the other hand, for all other configurations, tree-level Yukawa couplings $Qu^CH_u$ can be written using the quantum number assignments \eqref{eq:config_up_yukawa}.

We therefore need a minimum number of four stacks of branes in order to satisfy our conditions for the Yukawa couplings \eqref{eq:Yukawa_conditions} as well as the requirement \eqref{intersection_condition} that $H_d$ lives in a different intersection than the lepton doublet $L$. With the hypercharge embedding \eqref{eq:Madrid_model}, we find two distinct models fulfilling these conditions, whose spectrum is given in \eqref{eq:4_stacks_model_1} and \eqref{eq:4_stacks_model_2}. Analogous analysis for the seven other possible hypercharge embeddings listed in \cite{Anastasopoulos:2006da} can easily be carried out, leading to similar conclusions.


\vspace*{0.3cm}

\section*{Acknowledgements}
\label{ACKNOWL}
\addcontentsline{toc}{section}{Acknowledgements}

We would like to thank I. Antoniadis, K. Kaneta, E. Kiritsis, Y. Mambrini and J. Khlifi for valuable discussions.
P.A. was supported by FWF Austrian Science Fund via the SAP P30531-N27. F.R. was partially supported by the Cyprus R.I.F. Excellence hub/0421/0362 grant.

\newpage



\appendix
\numberwithin{equation}{section}
\section{Fermionic and bosonic modes on intersections}\label{app:a}

The following appendix is based on our previous work \cite{ANASTASOPOULOS2022115749} and is cited here for completeness. More details can be found there.
In order to perform an explicit calculation, we need to specify the details of the considered setup. The D-brane construction is based on three different stacks of D-branes. More precisely, it consists of a $D_a$-brane, a $D_b$-brane and a $D_c$-brane, which are wrapped and intersect each other non-trivially on a factorizable $6$-torus $\mathbb{T}^6=\mathbb{T}^2_1 \times \mathbb{T}^2_2 \times \mathbb{T}^2_3$. 
Such a D-brane model gives rise to the following intersection angles $\theta_i$, with $a_i \equiv\frac{\theta_i}{\pi}$:
\bea
&&a^1_{ab}  > 0,  \qquad \qquad  a^2_{ab}  >0,  \qquad \qquad  a^3_{ab}  < 0,  \qquad \qquad \sum a^i_{ab} =0 ~.\nn \\
&&a^1_{bc}  > 0,  \qquad \qquad  a^2_{bc}  > 0,  \qquad \qquad  a^3_{bc}  <0,  \qquad \qquad ~ \sum a^i_{bc} =0 ~. \nn\\
&&a^1_{ca}  < 0,  \qquad \qquad  a^2_{ca}  < 0,  \qquad \qquad  a^3_{ca}  < 0,  \qquad \qquad \sum a^i_{ca} =-2 ~.
\label{eq choice of angles}\eea
At each intersection, a massless fermion appears, which, in the case of preserved supersymmetry, is accompanied by a massless scalar corresponding to a four-dimensional superpartner. In order to guarantee and to provide $\mathcal{N}=1$ supersymmetry, the angles have to satisfy the triangle relations 
\bea
&& a^1_{ab} + a^1_{bc} +a^1_{ca} =0 ~~,~~~~  
a^2_{ab} + a^2_{bc} +a^2_{ca} =0 ~~,~~~~
a^3_{ab} + a^3_{bc} +a^3_{ca} =-2  ~~.~~~~
\label{concrete setup 2}
\eea
Furthermore, we find massless matter at each intersection and an entire tower of massive copies, whose mass scales with the intersection angle. These excitations are referred to as light stringy states. In scenarios with low string tension and small intersection angles, such states can be relatively light and potentially observed at the Large Hadron Collider (LHC) or in future experiments. 


\subsection*{Scalars at angles}
In the following, we focus on the intersections in the $ab$ sector between the $D_a$-brane and the $D_b$-brane. The angles must satisfy the conventions from (\ref{eq choice of angles}). In particular, this means that two intersection angles are positive, while the last one takes a negative value. The NS vacuum consists of a single massless state, which reads
\bea
\Phi(k)~:~~~~~
\psi_{-\frac{1}{2}-a^3_{ab}} \;|a^1_{ab},a^2_{ab},a^3_{ab}\rangle_{\mathrm{NS}},\quad \text{with} \quad \alpha^{\prime} m^{2}=\frac{1}{2}\sum_{i}^{3} a_{a b}^{i}=0  ~.
\eea
The associated mass squared operator vanishes in that case.
The vertex operator (VO) of this massless state in the canonical $(-1)$ super-ghost picture is given by
\bea
 V_{\Phi}^{(-1)}=g_{\Phi}\left[\Lambda_{a b}\right]_{\alpha}^{\beta} \Phi(x) e^{-\phi} \prod_{I=1}^{2} \sigma^{+}_{a_{a b}^{I}} e^{i a_{a b}^{I} H^{I}} \sigma^{+}_{1+a_{a b}^{3}} e^{i\left(1+a_{a b}^{3}\right) H^{3}} e^{i k X} ~,
\eea
where for the internal space $\mathbb{T}^{6}$ we get contributions from the bosonic twist fields $\sigma^{+}_{a}$ and the bosonized fermionic twist fields $e^{ia_I H_I}$. These twist fields incorporate the mixed boundary conditions of the open string stretched between intersecting branes. The additional $e^{ikX}$ comes from the four-dimensional spacetime structure, where the string can move freely. The  Chan-Paton factors $[\Lambda_{ab}]$ indicate that the oriented, open string is stretched between the two D-brane stacks $a$ and $b$. On each stack of D-branes, there lives a gauge group; thus the indices $\a$ and $\b$ run from one to the dimension of the fundamental representation of that gauge group. The string vertex coupling is denoted by $g_{\Phi}$. 

BRST symmetry requires that a Vertex operator obey the physical state condition $[Q_{BRST},V]=0$. Fulfilling this condition gives a double pole which vanishes for $\a' k^2=0$. 

Assuming that the angle  $a^1_{ab}$ is smaller than the rest, the lightest stringy states with masses $\a'\,m^2=a^1_{ab}$ include
\bea
&&\widetilde \Phi_1(k)~:~~~~~~~~~~~~~~~~~~
a_{a^1_{ab}} \psi_{-\frac{1}{2}-a^3_{ab}} |a^1_{ab},a^2_{ab},a^3_{ab}\rangle_{\mathrm{NS}}~,  ~~~~~~~  \\
&&\widetilde \Phi_2(k)~:~~~~~~~~~~~~~~~~~~~~~~~
\psi_{-\frac{1}{2}+a^2_{ab}} |a^1_{ab},a^2_{ab},a^3_{ab}\rangle_{\mathrm{NS}} ~.
\eea
The VO for these states are
\bea
V^{(-1)}_{\widetilde \Phi_1} &= g_\Phi [\Lambda_{ab}]^{\beta}_{\a} \, \widetilde \Phi_1(x) \, e^{-\phi} \;\tau_{a^1_{ab}} \,e^{i a^1_{ab} H_1}\; \sigma_{a^2_{ab}} \,e^{i a^2_{ab} H_2} 
 \;\sigma_{1+a^3_{ab}} \,e^{i \left(1+a^3_{ab}\right) H_3}\,e^{ipX} ~, \label{phi1}\\
V^{(-1)}_{\widetilde \Phi_2} &= g_\Phi [\Lambda_{ab}]^{\beta}_{\a} \, \widetilde \Phi_2(x) \, e^{-\phi}\; \sigma_{a^1_{ab}} \,e^{i a^1_{ab} H_1} 
 \;\sigma_{a^2_{ab}} \,e^{-i \left(1-a^2_{ab}\right) H_2}
\; \sigma_{1+a^3_{ab}} \,e^{i a^3_{ab} H_3}\, 
e^{ipX} ~.
\eea
Considering the BRST invariance of the VO's, a double pole appears, which vanishes if
\bea
\a' p^2 + a^1_{ab}=0
\label{eq:lightscalarmass}
\eea
for both VO's. Equation~(\ref{eq:lightscalarmass}) confirms that $\tilde\Phi_1$ and  $\tilde\Phi_2$ are massive scalars with mass square $a^1_{ab}/\a'$.
Here, we should notice that the single pole vanishes for both VO's.


\subsection*{Fermions at angles}

The other two states involved in our computations are two massless fermions from the Ramond sector. These two states are located at the intersections of the $D_b$-brane and $D_c$-brane as well as $D_c$-brane and $D_a$-brane. The two ground states are
\bea
\psi(k) ~:~~~~~
|\,a^1_{bc},a^2_{bc},a^3_{bc}\rangle_{\mathrm{R}} \label{eq vo phi}
~~~~~~~~~\text{and}~~~~~~~~~~ 
\chi(k) ~:~~~~~
|\,a^1_{ca},a^2_{ca},a^3_{ca}\rangle_{\mathrm{R}} \label{eq vo psi} ~.
\eea
Their associated VO's in the canonical $(-1/2)$ super-ghost picture is
\bea
V^{(-\frac{1}{2})}_{\psi} &=& g_{\psi}[\Lambda_{bc}]^\beta_\gamma ~ \psi^{\alpha}_i \, e^{-\phi/2} S_{\alpha}  
\label{VOpsi}\\
&& \times\;\sigma_{a^1_{bc}} \,e^{i \left(a^1_{bc} -\frac{1}{2}\right) H_1} \;
\; \sigma_{a^2_{bc}} \,e^{i \left(a^2_{bc} -\frac{1}{2}\right) H_2} \;
\; \sigma_{1+a^3_{bc}} \,e^{i \left(a^3_{bc} +\frac{1}{2}\right) H_3} \;
\, e^{ikX} ~,~~~~~~~~\nn\\
V^{(-\frac{1}{2})}_{\chi} &=& g_{\chi} [\Lambda_{ca}]^\a_\beta ~ \chi^{\alpha}_i \, e^{-\phi/2} S_{\alpha}  
\label{VOchi} \\
&& \times\;\sigma_{1+a^1_{ca}} \,e^{i \left(a^1_{ca} +\frac{1}{2}\right) H_1}\; \sigma_{1+a^2_{ca}} \,e^{i \left(a^2_{ca} +\frac{1}{2}\right) H_2}\;\sigma_{1+a^3_{ca}}  e^{i \left(a^3_{ca} +\frac{1}{2}\right) H_3}\;
\, e^{ikX} ~.~~~~~~~~\nn\eea
Apart from the spinor wave functions $\psi^{\alpha}_i$ and $\chi^{\alpha}_i$, we have an additional new type of field $S_\alpha$, which denotes a  $SO(1,3)$ spin field determined by the GSO projection. 

The mass squared operator vanishes for the spacetime fermions $\psi$ and $\chi$. Moreover, $\alpha^{\prime}m^2=0$ is independent of the choice of the angles. The physical state condition $[Q_{B R S T}, V_{\psi,\chi}^{(-1 / 2)}]=0$ yields a double and simple pole.
\begin{itemize}
    \item  The simple pole vanishes if we demand the equation of motion for a massless Weyl fermion, i.e.
    \begin{equation}
        k^\mu \bar\sigma^{\dot a a}_\mu \psi_a(k)= 0 \quad \text{and} \quad k^\mu \bar\sigma^{\dot a a}_\mu \chi_a(k)= 0 ~.
    \end{equation}
     \item  The double pole vanishes for $\alpha^{\prime}k^2=0$.
\end{itemize}

\providecommand{\href}[2]{#2}\begingroup\raggedright\endgroup


\end{document}